\documentclass[aps,preprint]{revtex4}%
\usepackage{amsfonts}
\usepackage{amsmath}
\usepackage{amssymb}
\usepackage{graphicx}%
\providecommand{\U}[1]{\protect\rule{.1in}{.1in}}

\begin{document}
\title{A new technique for the characterization of viscoelastic materials: theory,
experiments and comparison with DMA. }
\author{Elena Pierro$^{1}$, Giuseppe Carbone$^{2,3}$}
\affiliation{$^{1}$Scuola di Ingegneria, Universit\`{a} degli Studi della Basilicata, 85100
Potenza, Italy}
\affiliation{$^{2}$Department of Mechanics, Mathematics and Management, Polytechnic
University of Bari, V.le Japigia, 182, 70126, Bari, Italy}
\affiliation{$^{3}$Physics Department M. Merlin, CNR Institute for Photonics and
Nanotechnologies U.O.S. Bari, via Amendola 173, Bari, 70126, Italy}
\begin{abstract}
In this paper we present a theoretical and experimental study aimed at
characterizing the hysteretic properties of viscoelastic materials. In the
last decades viscoelastic materials have become a reference for new
technological applications, which require lightweight, deformable but
ultra-tough structures. The need to have a complete and precise knowledge of
their mechanical properties, hence, is of utmost importance. The presented
study is focused on the dynamics of a viscoelastic beam, which is both
experimentally investigated and theoretically characterized by means of an
accurate analytical model. In this way it is possible to fit the experimental
curves to determine the complex modulus. Our proposed approach enables the
optimal fitting of the viscoelastic modulus of the material by using the
appropriate number of relaxation times, on the basis of the frequency range
considered. Moreover, by varying the length of the beams, the frequency range
of interest can be changed/enlarged. Our results are tested against those
obtained with a well established and reliable technique as compared with
experimental results from the Dynamic Mechanical Analysis (DMA), thus
definitively establishing the feasibility, accuracy and reliability of the
presented technique.

\end{abstract}
\startpage{1}
\endpage{2}
\maketitle

\part{Introduction}

Recent scientific advancements in field of automotive, electronics,
micromechanical systems, pipe technologies, have led to new technologies where
the use of lightweight, tough, soft and high deformable materials has become
ubiquitous. In this scenario, viscoelastic materials have spread in many
different contexts, from seals \cite{Bottiglione2009} to bio-inspired
adhesives \cite{Carbone2011,Carbone2012,Carbone2012bis,Carbone2013bis},
because of their superior damping and frictional properties. For an
appropriate use of such materials, however, the proper knowledge of their
mechanical properties is a basic requirement. Along this line, the most
popular technique to characterize the viscoelastic modulus of such materials
is the Dynamic Mechanical Analysis (DMA) \cite{Chartoff2009,Huayamares2020},
which allows the measurement of the viscoelastic complex modulus depending on
both frequency and temperature. In particular, it consists of imposing a small
cyclic strain on a sample and measuring the resulting stress response, or
equivalently, imposing a cyclic stress on a sample and measuring the resultant
strain response. Nevertheless, such an experimental procedure exhibits
different limits (e.g. high frequency characterization is considerably
difficult \cite{Nolle1948,Nijenhuis1980,Esmaeeli2019}) and requires expensive
test equipment. In this view, several techniques have been proposed, based on
the vibrational response of beam like structures. In Ref. \cite{Pritz1982},
the complex modulus of acoustic materials using a transfer function method of
a lumped mechanical model was utilized. In Ref. \cite{Trendafilova1994} the
response of the endpoint of an impacted beam was measured in terms of
displacements, by means of electro-optical transducers, and then an iterative
numerical scheme was considered to retrieve the viscoelastic modulus. Other
experimental procedures have been recently presented, with the aim of
simplifying the setup, as in Ref. \cite{Casimir2012} where a double pendulum
was utilized to excite a viscoelastic sample without any other source of
external excitation, and recorded oscillations were induced by gravity. In
Ref. \cite{Cortes2007} a cantilever beam was excited by means of a seismic
force, and curve fitting of experimental data with a fractional derivative
model was employed to characterize the complex modulus. However, in all the
presented works dealing with mechanical characterization of viscoelastic
materials, there is no the simultaneous presence of i) a very simple setup,
ii) an analytical model to describe the dynamics of the vibrational system
considered, and iii) a constitutive model able to accurately capture the
behaviour of viscoelastic materials in a wide frequency range. Moving from
these facts, in this paper we present a rigorous easy-to-use approach for
determining the viscoelastic modulus, based on the experimental vibrational
identification of viscoelastic beams with different lengths. Both a very
simple setup is utilized for acquisitions, and an accurate analytical model of
the beam are considered to determine the viscoelastic modulus, which takes
into account multiple relaxation times of the material. In particular, by
properly changing the length of the considered beam, it is possible to broaden
the frequency range under analysis, and by selecting the appropriate number of
relaxation times it is possible to optimize the fitting procedure. The results
presented in this paper show that these two aspects are of pivotal importance
to correctly determine the viscoelastic complex modulus, and they open new
paths towards challenging further improvements. The paper is organized as
follows: at first, the analytical model of a viscoelastic beam dynamics is
recalled, then the experimental setup and the data acquisition are explained
in detail. Finally, the curve fitting scheme is defined and results are
discussed in depth, with particular emphasis on the comparison with DMA results.

\section{Theoretical model of the viscoelastic beam dynamics}

In this section, we derive the analytical formulation of the viscoelastic beam
vibrational response. The main purpose is to get a simple-to-use formula,
which can be utilized to characterize the viscoelastic modulus, by fitting the
experimental acquisitions. It is known, in particular, that for viscoelastic
materials, the stress-strain relation is governed by the following integral
\cite{Christensen}
\begin{equation}
\sigma\left(  x,t\right)  =\int_{-\infty}^{t}G\left(  t-\tau\right)
\dot{\varepsilon}\left(  x,\tau\right)  \mathrm{d}\tau\label{stress-strain}%
\end{equation}
where $\dot{\varepsilon}(t)$ is time derivative of the strain, $\sigma(t)$ is
the stress, $G\left(  t-\tau\right)  $ is the so called relaxation function.
The viscoelastic complex modulus $E\left(  s\right)  $ is closely related to
the relaxation function $G\left(  t\right)  $, and the simple equality
$E\left(  s\right)  =sG\left(  s\right)  $ exists in the Laplace domain. In
this domain, in particular, it is possible to represent the complex modulus
$E\left(  s\right)  $ as the following series%
\begin{equation}
E\left(  s\right)  =E_{0}+\sum_{k}E_{k}\frac{s\tau_{k}}{1+s\tau_{k}}
\label{ElasticModulusLaplace}%
\end{equation}
which derives from the generalized Maxwell model, consisting of a spring with
elastic constant $E_{0}$, that represents the elastic modulus of the material
at zero-frequency, and $k$ Maxwell elements connected in parallel, i.e. spring
elements characterized by both the relaxation time $\tau_{k}$ and the elastic
modulus $E_{k}$. By considering the above Eqs.(\ref{stress-strain}%
)-(\ref{ElasticModulusLaplace}) in the theoretical model utilized to fit the
experimental responses of the vibrating beam, it is possible to establish the
optimal number of relaxation times to better capture the viscoelastic
behaviour in a certain frequency range.

\begin{figure}[ptb]
\begin{center}
\includegraphics[
height=9cm,
]{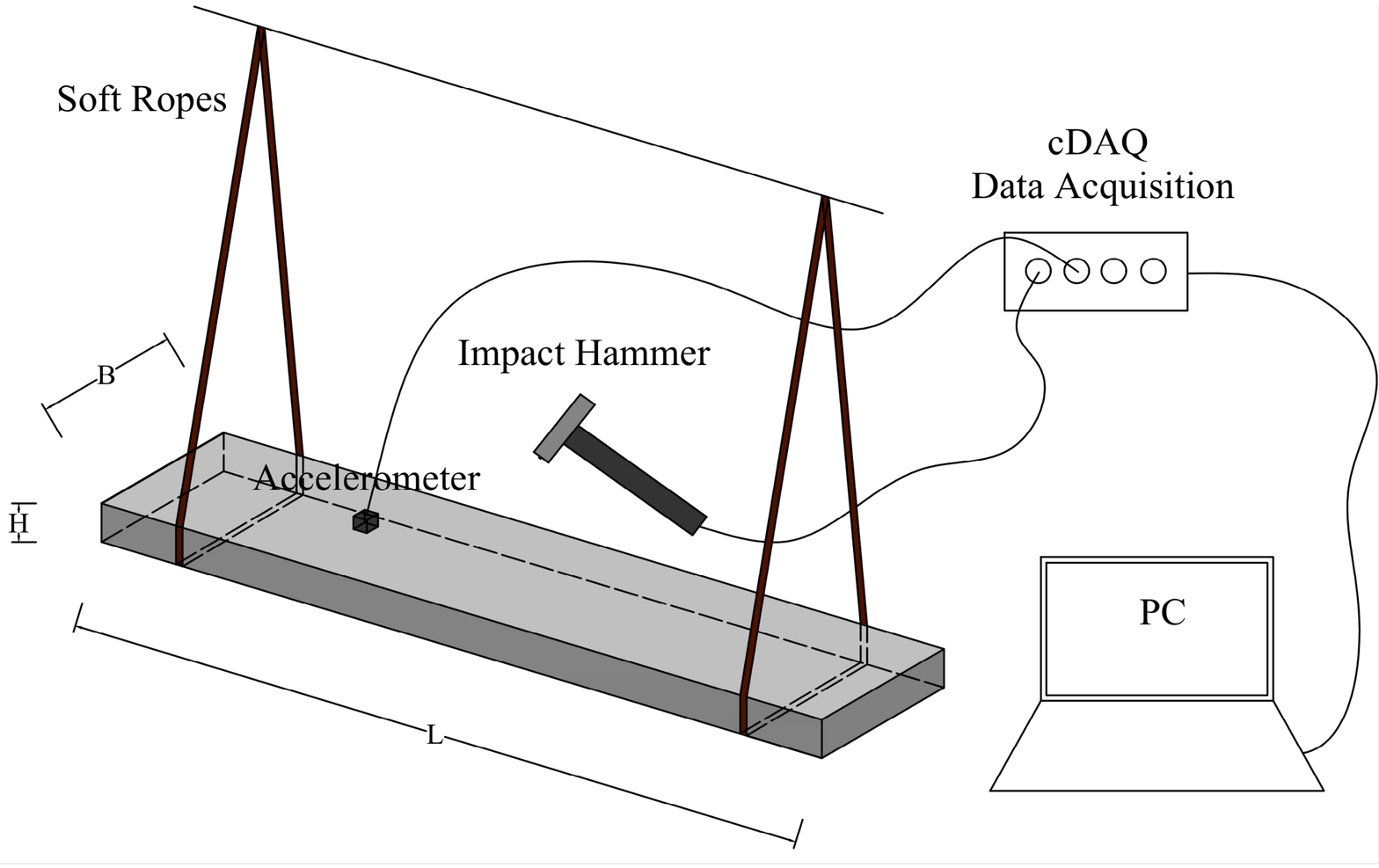}
\end{center}
\caption{The schematic of the test rig.}%
\label{Figure 1}%
\end{figure}

The geometrical characteristics of the beam considered in the present paper
are chosen in order to follow the Bernoulli theory of flexural vibrations,
i.e. $L\gg W$, $L\gg B$, being $L$ the length of the beam, $W$ and $B$
respectively the width and\ the thickness of the rectangular cross section.

\begin{figure}[ptb]
\centering
\includegraphics[
height=9cm,
]{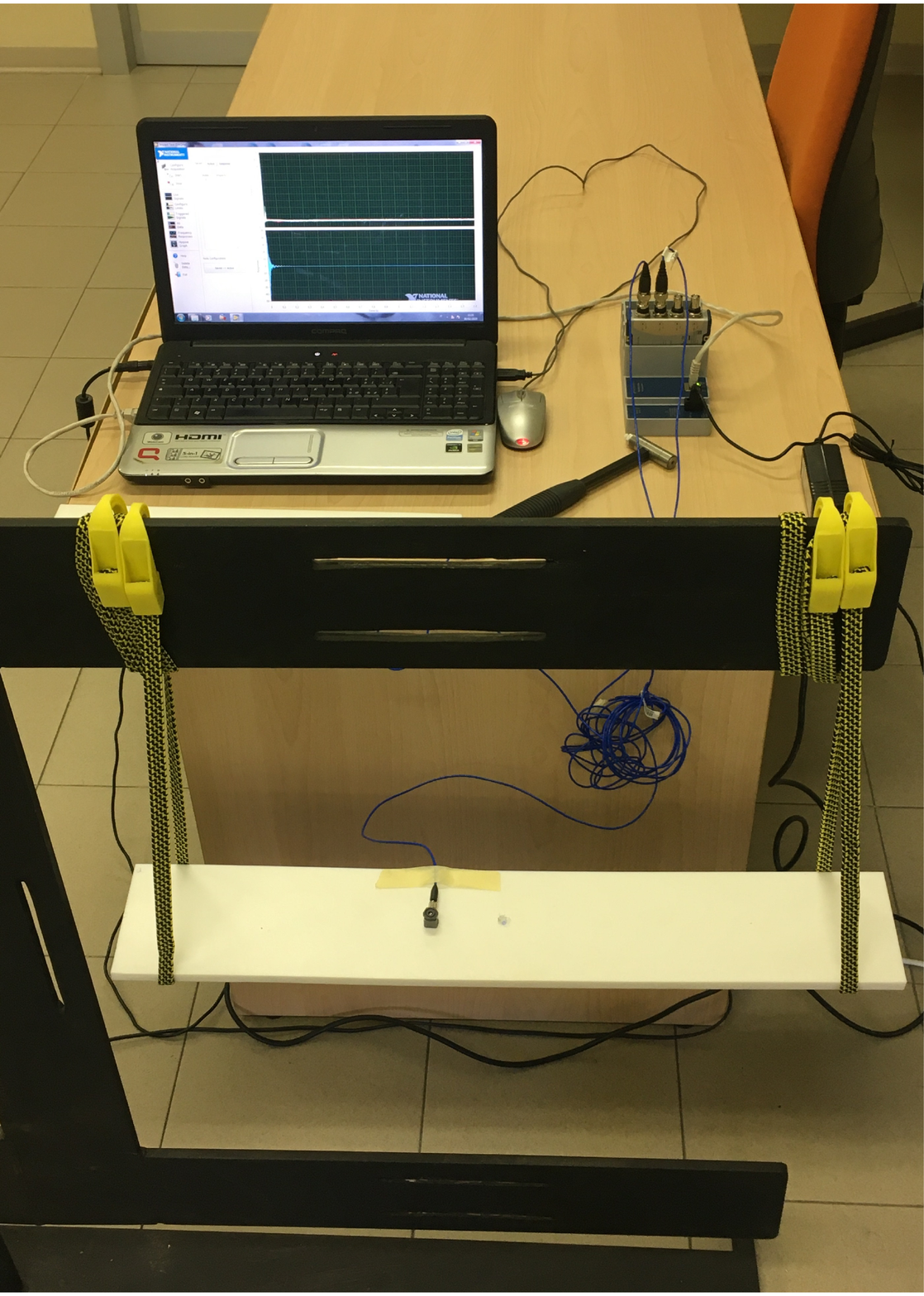} \caption{The experimental setup, consisting of i) the
suspended viscoelastic beam (length $L=60$ [cm]) with a PCB 333B30 ICP
accelerometer, glued on the upper surface, ii) the cDAQ-9184 NI data
acquiring, iii) the PCB 086C03 ICP Impact Hammer and iv) a portable pc.}%
\label{Figure 2}%
\end{figure}In particular, by assuming that the transversal displacement
$\left\vert u\left(  x,t\right)  \right\vert \ll L$, it is possible to neglect
the contribute of shear stress, which is always very low at the first
resonances. The equation of motion can be therefore written as
\cite{Inman1996}%
\begin{equation}
J_{xz}\int_{-\infty}^{t}E\left(  t-\tau\right)  \frac{\partial^{4}u\left(
x,\tau\right)  }{\partial x^{4}}\mathrm{d}\tau+\mu~\frac{\partial^{2}u\left(
x,t\right)  }{\partial t^{2}}=f\left(  x,t\right)  \label{motionEq}%
\end{equation}
being $\mu=\rho A$, $\rho$ the bulk density of the material, $A=WB$ the cross
section area, $J_{xz}=(1/12)WB^{3}$ the moment of intertia, and $f\left(
x,t\right)  $ the generic force acting on the beam. The solution of
Eq.(\ref{motionEq}) can be formulated in terms of the eigenfunctions $\phi
_{n}\left(  x\right)  $ by means of the decomposition of the system response%
\begin{equation}
u\left(  x,t\right)  =\sum_{n=1}^{+\infty}\phi_{n}\left(  x\right)
q_{n}\left(  t\right)  \label{sepvariable}%
\end{equation}
\begin{figure}[ptb]
\begin{center}
\includegraphics[
height=6cm,
]{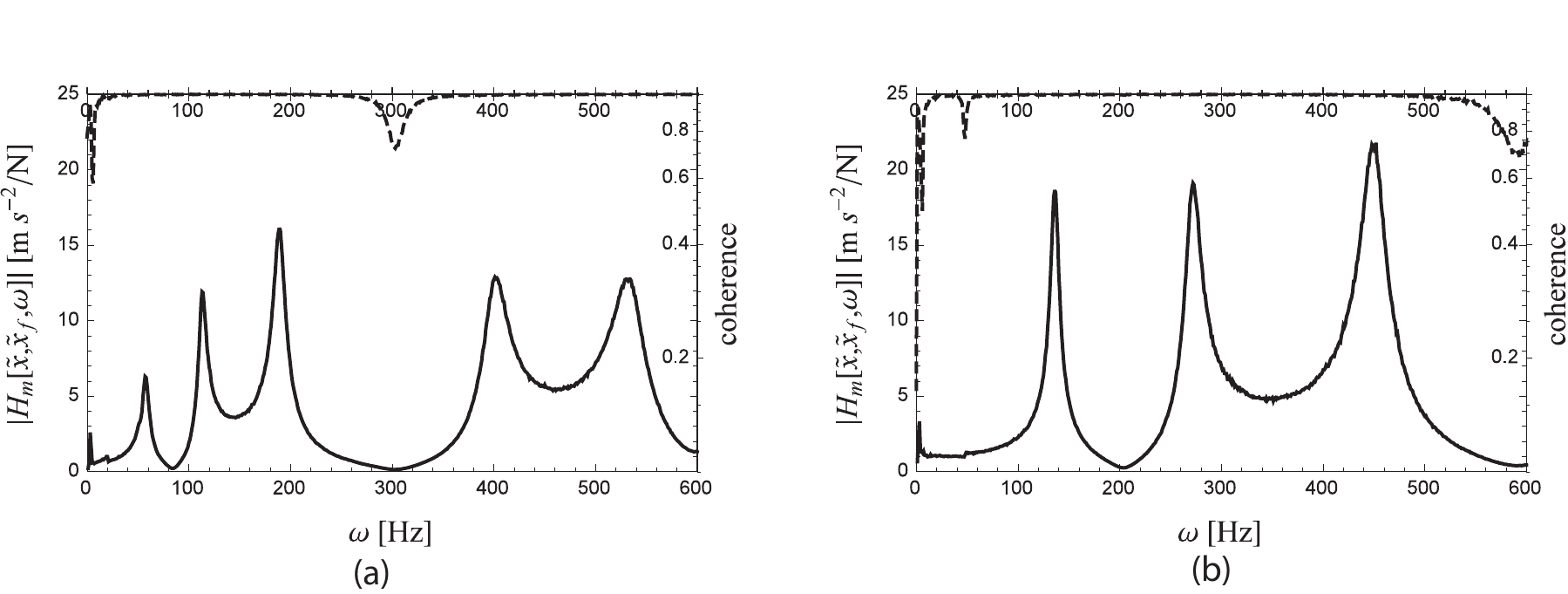}
\end{center}
\caption{The measured frequency response functions $H_{m}\left(  \tilde
{x},\tilde{x}_{f},\omega\right)  $ (solid lines) and the coherence (dashed
lines), in the range $0-600$ [Hz], for the beams of length $L=60$ [cm] (a) and
$L=40$ [cm] (b).}%
\label{Figure 3}%
\end{figure}It can be shown that the eigenfunctions $\phi_{n}\left(  x\right)
$ do not change if we consider a viscoelastic material instead of a perfectly
elastic one. The eigenfunctions $\phi_{n}\left(  x\right)  $, in particular,
can be calculated by solving the homogeneous problem
\begin{equation}
J_{xz}\int_{-\infty}^{t}E\left(  t-\tau\right)  \frac{\partial^{4}u\left(
x,\tau\right)  }{\partial x^{4}}\mathrm{d}\tau+\mu~\frac{\partial^{2}u\left(
x,t\right)  }{\partial t^{2}}=0 \label{autoproblem}%
\end{equation}
with the opportune boundary conditions. In the Laplace domain, by considering
the initial conditions equal to zero, Eq.(\ref{autoproblem}) becomes%
\begin{equation}
\phi_{xxxx}\left(  x,s\right)  -\beta_{eq}^{4}\left(  s\right)  \phi\left(
x,s\right)  =0 \label{autoproblemLaplace}%
\end{equation}
($\phi_{x}\left(  x,s\right)  =\partial\phi\left(  x,s\right)  /\partial x$)
having defined\begin{figure}[ptb]
\begin{center}
\includegraphics[
height=5cm,
]{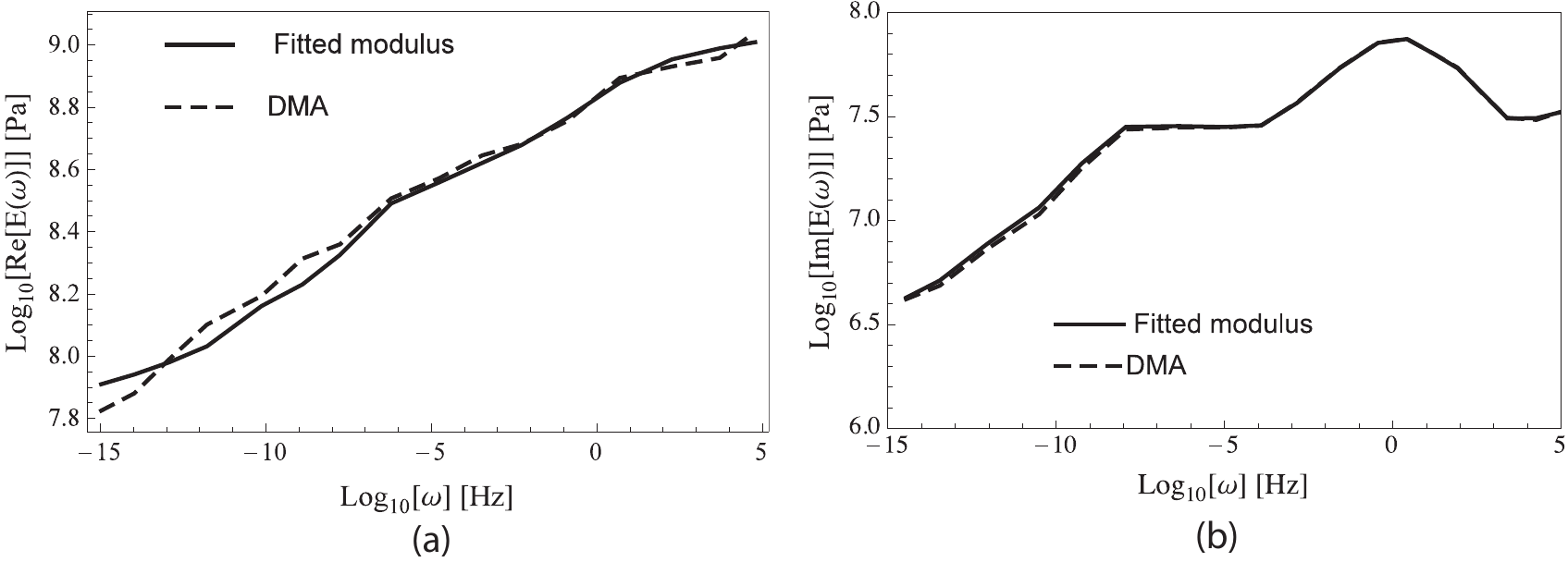}
\end{center}
\caption{Master curves of LUBRIFLON$^{\textregistered}$ (Dixon Resine
\cite{Dixon}) at 20${{}^{\circ}}$C (dashed lines), fitted by means of
Eq.(\ref{ElasticModulusLaplace}) by considering $50$ relaxation times (solid
lines), as real part $\operatorname{Re}[E\left(  \omega\right)  ]$ (a) and
imaginary part $\operatorname{Im}[E\left(  \omega\right)  ]$ (b) of the
viscoelastic modulus $E\left(  \omega\right)  $.}%
\label{Figure 11}%
\end{figure}%
\begin{equation}
\beta_{eq}^{4}\left(  s\right)  =-\frac{\mu~s^{2}}{J_{xz}E\left(  s\right)
}=-\frac{\mu~s^{2}}{J_{xz}}C\left(  s\right)  \label{beta_equivalent}%
\end{equation}
with $C\left(  s\right)  =1/E\left(  s\right)  $ the compliance of the
viscoelastic material. In the present study, we determine experimentally the
beam response when it is suspended at a fixed frame, in the so called
"free-free" boundary condition. From a mathematical point of view, this set-up
corresponds to the following mathematical conditions%
\begin{align}
\phi_{xx}\left(  0,s\right)   &  =0\label{BC autofunction}\\
\phi_{xxx}\left(  0,s\right)   &  =0\nonumber\\
\phi_{xx}\left(  L,s\right)   &  =0\nonumber\\
\phi_{xxx}\left(  L,s\right)   &  =0\nonumber
\end{align}
\begin{figure}[ptb]
\begin{center}
\includegraphics[
height=4cm,
]{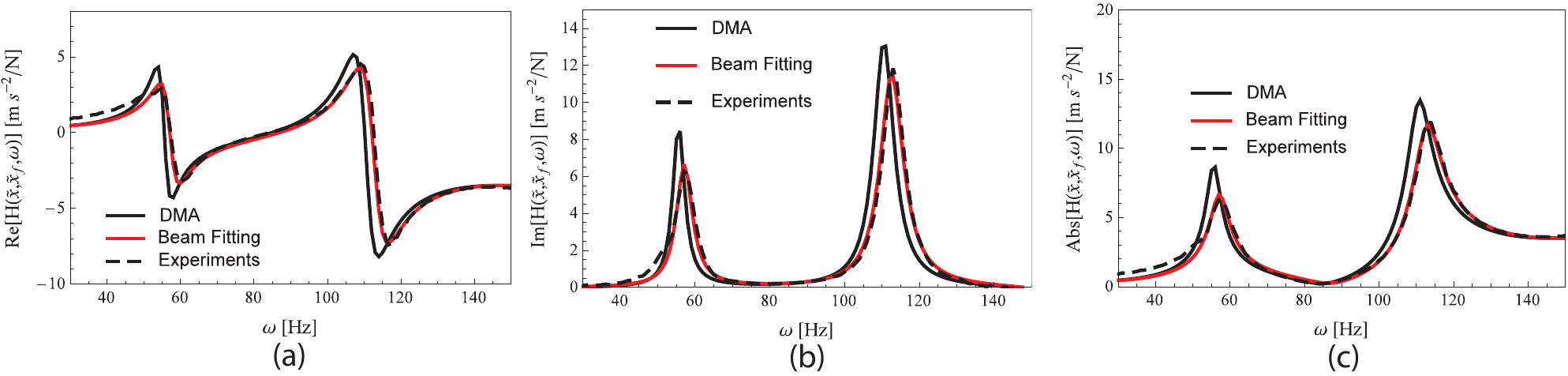}
\end{center}
\caption{The comparison between the measured FRF $H_{m}\left(  \tilde
{x},\tilde{x}_{f},\omega\right)  $ (black dashed lines) and the theoretical
FRF $H_{th}\left(  \tilde{x},\tilde{x}_{f},\omega\right)  $, obtained through
beam-best fitting (red lines) and DMA data (black solid lines), for the beam
of length $L=60$ [cm]. Curves are shown in terms of real part (a), imaginary
part (b) and absolute value (c) of the FRFs.}%
\label{Figure 4}%
\end{figure}From the solution of Eq.(\ref{autoproblemLaplace}), which is of
the following type%
\begin{equation}
\phi(x,s)=W_{1}\cos\left[  {\beta}_{eq}{\left(  s\right)  x}\right]
+W_{2}\sin\left[  {\beta}_{eq}{\left(  s\right)  x}\right]  +W_{3}\cosh\left[
{\beta}_{eq}{\left(  s\right)  x}\right]  +W_{4}\sinh\left[  {\beta}%
_{eq}{\left(  s\right)  x}\right]
\end{equation}
it is possible to derive the well known equation%
\begin{equation}
\left[  1-\cos\left(  \beta_{eq}L\right)  \cosh\left(  \beta_{eq}L\right)
\right]  =0 \label{modes_eq}%
\end{equation}
by simply forcing equal to zero the determinant of the system matrix obtained
from Eqs.(\ref{BC autofunction}). Let us observe that the solutions $\beta
_{n}L=c_{n}$ of Eq.(\ref{modes_eq}) are the same of the perfectly elastic
case, and can be substituted in Eq.(\ref{beta_equivalent}) to calculate the
complex conjugate eigenvalues $s_{n}$ corresponding to the $n$ modes of the
beam, and the real poles $s_{k}$ related to the material viscoelasticity
\cite{Pierro2019,Pierro2020}. Furthermore, by means of the solutions
$\beta_{n}L=c_{n}$ of Eq.(\ref{modes_eq}), it is possible to derive the
following eigenfunctions $\phi_{n}\left(  x\right)  $
\begin{equation}
\phi_{n}\left(  x\right)  =\cosh\left(  \beta_{n}x\right)  +\cos\left(
\beta_{n}x\right)  -\frac{\cosh\left(  \beta_{n}L\right)  -\cos\left(
\beta_{n}L\right)  }{\sinh\left(  \beta_{n}L\right)  -\sin\left(  \beta
_{n}L\right)  }\left[  \sinh\left(  \beta_{n}x\right)  +\sin\left(  \beta
_{n}x\right)  \right]  \label{modes}%
\end{equation}
which have the same analytical form of the eigenfunctions of a beam made of an
elastic material. In particular, these functions follow the orthogonality
condition%
\begin{equation}
\frac{1}{L}\int_{0}^{L}\phi_{n}\left(  x\right)  \phi_{m}\left(  x\right)
\mathrm{d}x=\delta_{nm} \label{orthogonality}%
\end{equation}
being $\delta_{nm}$ the Kronecker delta function, as well as from
Eq.(\ref{autoproblemLaplace}) one gets\begin{figure}[ptb]
\begin{center}
\includegraphics[
height=5cm,
]{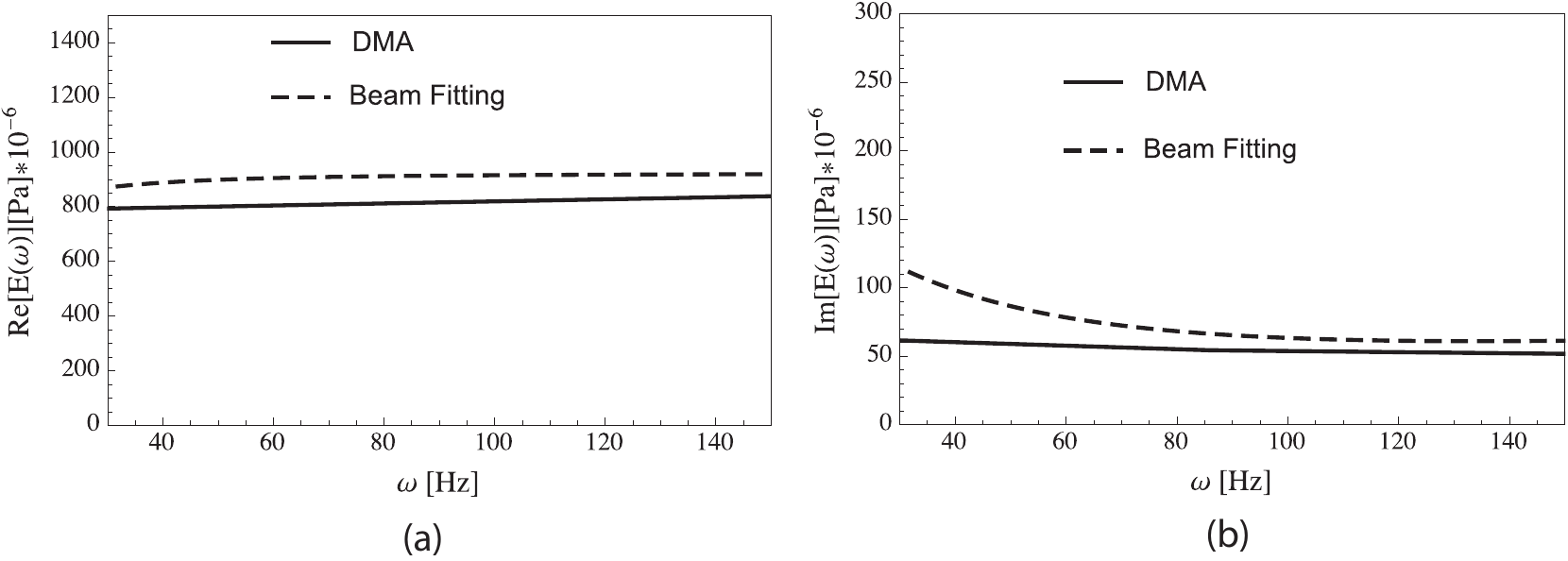}
\end{center}
\caption{The real part (a) and the imaginary part (b) of the viscoelastic
modulus $E\left(  \omega\right)  $, obtained by fitting the vibrational
response of the beam with $L=60$ [cm] (dashed line) and by DMA (solid line),
in the frequency range $30-150$ [Hz].}%
\label{Figure 5}%
\end{figure}%
\begin{equation}
\frac{1}{L}\int_{0}^{L}\left(  \phi_{n}\right)  _{xxxx}\left(  x\right)
\phi_{m}\left(  x\right)  \mathrm{d}x=\frac{1}{L}\int_{0}^{L}\phi_{n}\left(
x\right)  \beta_{n}^{4}\phi_{m}\left(  x\right)  \mathrm{d}x=\delta_{nm}%
\beta_{n}^{4} \label{orthog_der}%
\end{equation}
Through the above Eqs.(\ref{orthogonality})-(\ref{orthog_der}), and by
defining the projected solution $u_{m}\left(  t\right)  $ on the $m_{th}$
eigenfunction $\phi_{m}\left(  x\right)  $ as%

\begin{equation}
u_{m}\left(  t\right)  =\left\langle u\left(  x,t\right)  \phi_{m}\left(
x\right)  \right\rangle =\frac{1}{L}\int_{0}^{L}u\left(  x,t\right)  \phi
_{m}\left(  x\right)  \mathrm{d}x \label{sol_projected}%
\end{equation}
it is possible to rewrite Eq.(\ref{motionEq}), after simple calculations, as
following%
\begin{equation}
\mu\ddot{q}_{n}\left(  t\right)  +J_{xz}\beta_{n}^{4}\int_{-\infty}%
^{t}E\left(  t-\tau\right)  q_{n}\left(  \tau\right)  \mathrm{d}\tau
=f_{n}\left(  t\right)  \label{Eq projected time}%
\end{equation}
where $f_{n}\left(  t\right)  =\frac{1}{L}\int_{0}^{L}f\left(  x,t\right)
\phi_{n}\left(  x\right)  \mathrm{d}x$ is the projected force. The Laplace
Transform of Eq.(\ref{Eq projected time}), with initial conditions equal to
zero, is\begin{figure}[ptb]
\begin{center}
\includegraphics[
height=4cm,
]{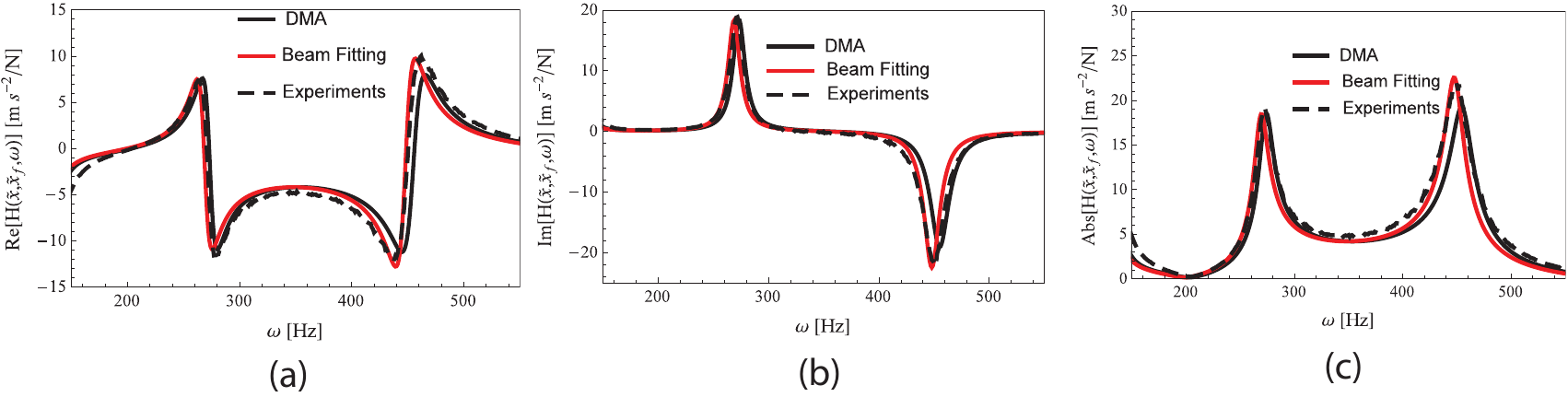}
\end{center}
\caption{The comparison between the measured FRF $H_{m}\left(  \tilde
{x},\tilde{x}_{f},\omega\right)  $ (black dashed lines) and the theoretical
FRF $H_{th}\left(  \tilde{x},\tilde{x}_{f},\omega\right)  $, obtained through
beam-best fitting (red lines) and DMA data (black solid lines), for the beam
of length $L=40$ [cm]. Curves are shown in terms of real part (a), imaginary
part (b) and absolute value (c) of the FRFs.}%
\label{Figure 6}%
\end{figure}%
\begin{equation}
\mu s^{2}Q_{n}\left(  s\right)  +J_{xz}\beta_{n}^{4}E\left(  s\right)
Q_{n}\left(  s\right)  =F_{n}\left(  s\right)  \label{Eq projected Laplace}%
\end{equation}
and therefore the system response, defined in Eq.(\ref{sepvariable}), becomes
in the Laplace domain
\begin{equation}
U\left(  x,s\right)  =\sum_{n=1}^{+\infty}\phi_{n}\left(  x\right)
Q_{n}\left(  s\right)  =\sum_{n=1}^{+\infty}\phi_{n}\left(  x\right)
\frac{F_{n}\left(  x,s\right)  }{\mu s^{2}+J_{xz}\beta_{n}^{4}E\left(
s\right)  } \label{LaplaceSolution}%
\end{equation}
For the scope of our investigation, we need to further modify
Eq.(\ref{LaplaceSolution}). Indeed, we experimentally excite the beam (see
Section II) by means of an impact hammer, in the section $x=x_{f}$, at the
instant $t=t_{0}$. Analytically, this condition is equivalent to consider as
forcing term, a Dirac Delta of constant amplitude $F_{0}$, in both the time
and the spatial domains $f\left(  x,t\right)  =F_{0}\delta\left(
x-x_{f}\right)  \delta\left(  t-t_{0}\right)  $, i.e. in the Laplace domain
the projected force is $F_{n}=\int_{0}^{L}F_{0}\delta\left(  x-x_{f}\right)
\phi_{n}\left(  x\right)  \mathrm{d}x=F_{0}\phi_{n}\left(  x_{f}\right)  $.
Hence, the analytic response of the beam can be rewritten
as\begin{figure}[ptb]
\begin{center}
\includegraphics[
height=5cm,
]{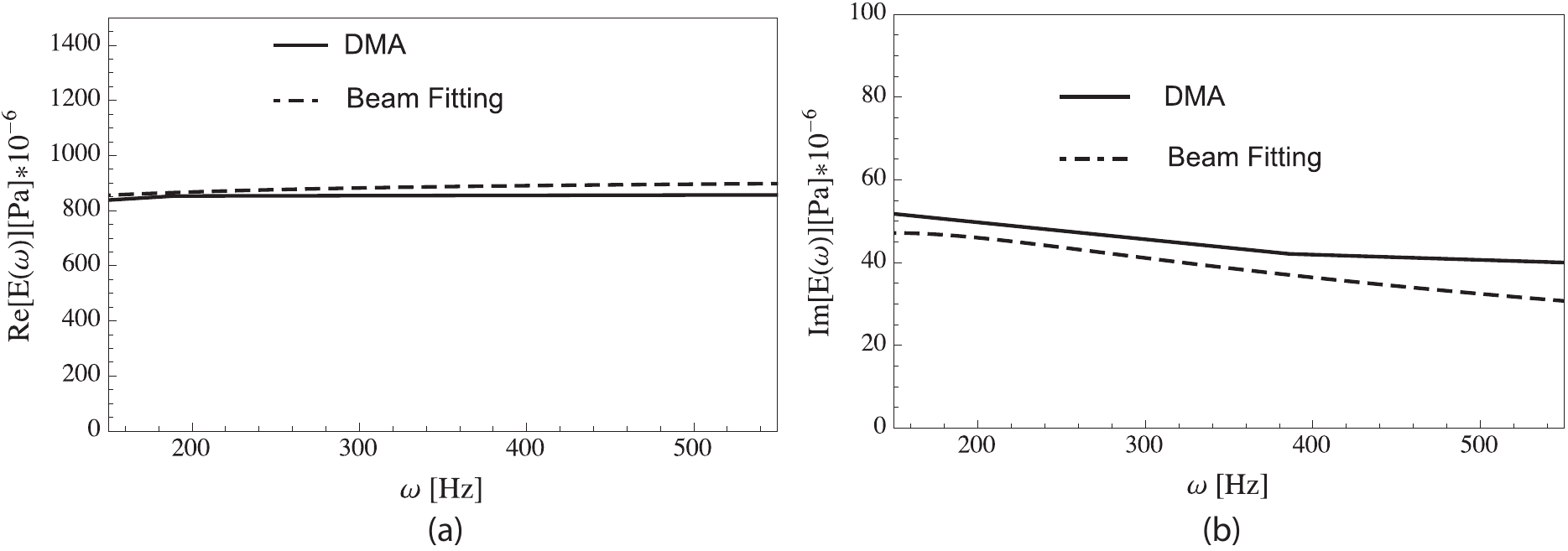}
\end{center}
\caption{The real part (a) and the imaginary part (b) of the viscoelastic
modulus $E\left(  \omega\right)  $, obtained by fitting the vibrational
response of the beam with $L=40$ [cm] (dashed line) and by DMA (solid line),
in the frequency range $150-550$ [Hz].}%
\label{Figure 7}%
\end{figure}%
\begin{equation}
U\left(  x,x_{f},s\right)  =F_{0}\sum_{n=1}^{+\infty}\frac{\phi_{n}\left(
x\right)  \phi_{n}\left(  x_{f}\right)  }{\mu s^{2}+J_{xz}\beta_{n}%
^{4}E\left(  s\right)  } \label{SemplifiedSystemResponse}%
\end{equation}
The beam response, derived in the above Eq.(\ref{SemplifiedSystemResponse}),
can be utilized to determine the viscoelastic modulus $E\left(  s\right)  $,
previously defined by Eq.(\ref{ElasticModulusLaplace}). In particular, the
following theoretical frequency response function (FRF), can be defined in
terms of inertance%
\begin{equation}
H_{th}\left(  x,x_{f},\mathrm{i}\omega\right)  =\frac{A\left(  x,\mathrm{i}%
\omega\right)  }{F_{0}}=(\mathrm{i}\omega)^{2}\sum_{n=1}^{+\infty}\frac
{\phi_{n}\left(  x\right)  \phi_{n}\left(  x_{f}\right)  }{\mu\left(
\mathrm{i}\omega\right)  ^{2}+J_{xz}\beta_{n}^{4}E\left(  \mathrm{i}%
\omega\right)  } \label{FRFTh}%
\end{equation}
being the acceleration $A\left(  x,\mathrm{i}\omega\right)  =U\left(
x,\mathrm{i}\omega\right)  (\mathrm{i}\omega)^{2}$. In this way,
Eq.(\ref{FRFTh}) can be utilized to fit the experimentally acquired FRF, as
discussed in the next Section.

\section{Experimental test}

Two viscoelastic beams made of LUBRIFLON$^{\textregistered}$ (Dixon Resine
\cite{Dixon}), with thickness $B=1$ [cm], width $W=10$ [cm], lengths $L=60$
[cm] and $L=40$ [cm], were suspended at a fixed frame through soft ropes.
Different lengths, in particular, enable to enlarge the frequency range of
interest, thus resulting in a better characterization of the material damping
properties, as it will be thoroughly discussed in the next Sections.
Furthermore, by considering beams with different lengths, it is possible to
survey potential peaks suppression or mitigation, according to the theoretical
studies previously presented in Ref.\cite{Pierro2019,Pierro2020}. The
schematic of the test rig is drawn in Figure \ref{Figure 1}. The basic
experimental setup (laboratory of Applied Mechanics, University of Basilicata,
Potenza, Italy), is shown in Figure \ref{Figure 2}. This kind of setup, which
represents the free-free boundary condition, is suitable in order to avoid
external influences on damping due to constraints \cite{Ewins1984}, as it
happens for example when the beam is clamped. The cDAQ-9184 CompactDAQ
(National Instruments) data acquiring has been utilized to collect the time
histories, through the NI Sound and Vibration Toolkit included in LabVIEW
(National Instruments). The slender beam has been excited in the $z$-direction
through the PCB 086C03 ICP Impact Hammer, and the accelerations have been
acquired, in the same direction, by means of a PCB 333B30 ICP Accelerometer.
\begin{figure}[ptb]
\begin{center}
\includegraphics[
height=5cm,
]{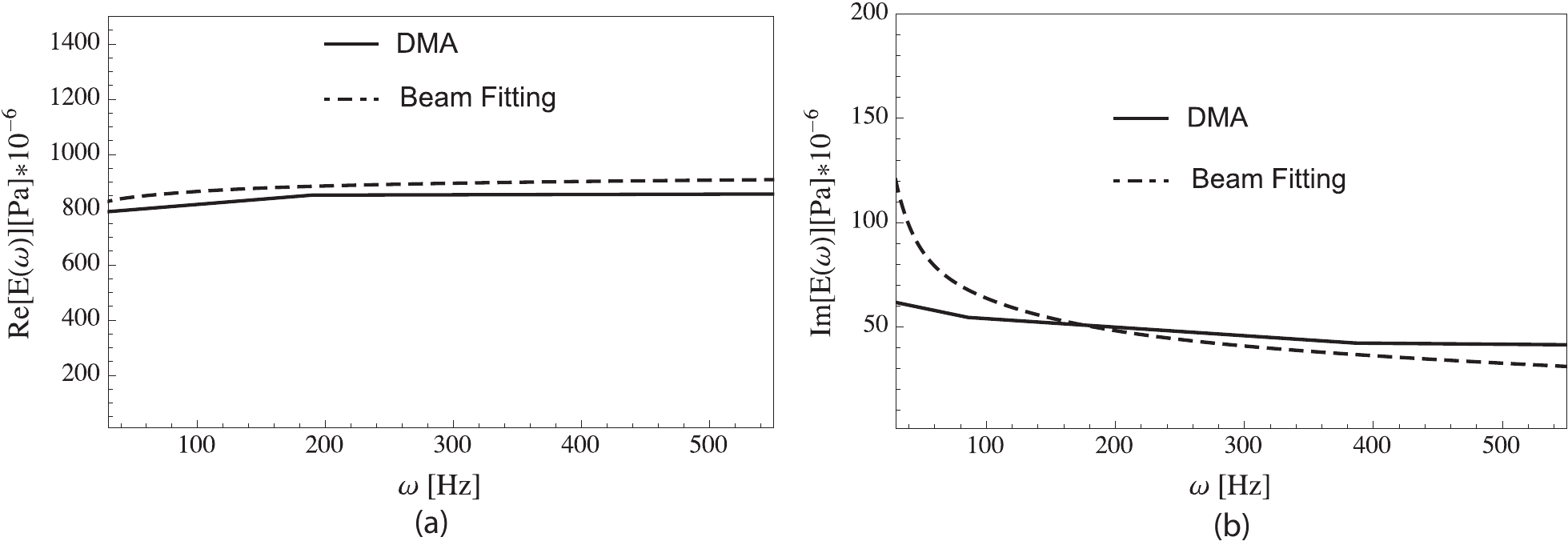}
\end{center}
\caption{The real part (a) and the imaginary part (b) of the viscoelastic
modulus $E\left(  \omega\right)  $, in the frequency range $30-550$ [Hz],
obtained from DMA (solid line) and by fitting the two moduli determined
through the vibrational responses of the two beams with different lengths
(dashed line).}%
\label{Figure 8}%
\end{figure}The beam sections chosen for the impacting excitation and for the
acceleration acquisitions were respectively $\tilde{x}_{f}=0.8L$ and
$\tilde{x}=0.4L$. The motivation behind this choice lies in the fact that, in
this way, by properly avoiding the nodal points, the first vibrational modes
$\phi_{1,5}\left(  x_{f}\right)  $ should all be present in the measures FRFs,
in the frequency range under analysis. However, it is expected a mitigation of
both the first and fifth peaks, since the section of the input force
$\tilde{x}_{f}$ is close to nodal points for the corresponding two mode shapes.

We acquired a group of $10$ time histories, each lasting $1$ [s], with
sampling frequency $f_{s}=25600$ [Hz]. It should be observed that, because of
the heavy damped material, the signal decreased to zero at about $1/5$ of the
acquisition time, so there was no need to apply any windows to time histories.
Then, we have calculated the Fast Fourier Transform (FFT) of the ten averaged
time histories, for both the accelerations $A\left(  \tilde{x},\omega\right)
$ and the impacting forces $F\left(  \tilde{x}_{f},\omega\right)  $. Finally,
the $H_{1}$ estimator \cite{Pintelon2001} has been considered to determine the
measured frequency response functions $H_{m}\left(  \tilde{x},\tilde{x}%
_{f},\omega\right)  $, which are shown in Figure \ref{Figure 3}, for the beams
of length $L=60$ [cm] (Figure \ref{Figure 3}-a) and $L=40$ [cm] (Figure
\ref{Figure 3}-b), in terms of absolute value of the function $H_{m}\left(
\tilde{x},\tilde{x}_{f},\omega\right)  $ (solid lines), in the frequency range
$0-600$ [Hz]. Moreover, the coherence function \cite{Pintelon2001} (dashed
lines) for each acquisition is shown, for the same frequency range. It is
possible to observe that, as expected, same resonances moves forward higher
frequencies, by decreasing the beam length $L$. However, in contrast to a
perfectly elastic beam, the amplitude of such peaks changes. In particular,
the second peak, which is at $\omega\simeq60$ [Hz] for the beam of length
$L=60$ [cm] (Figure \ref{Figure 3}-a), moves to $\omega\simeq120$ [Hz] for the
beam with $L=40$ [cm] (Figure \ref{Figure 3}-b) and its amplitude increases.
This circumstance suggests that the transition region of the material, where
damping effects are more significant, should be found at lower frequencies.
For both the material and the beam geometry considered in the present study, a
slight mitigation of the resonances can be observed, and not a complete peak
suppression \cite{Pierro2019,Pierro2020}. This fact helps in interpreting the
nature of the peaks in the frequency range considered, and thus enables us to
perform a correct viscoelastic modulus fitting.

\section{Viscoelastic Parameters Identification}

By observing the coherence functions in Figure \ref{Figure 3}, it is possible
to notice that some problems occurred before the second and after the fourth
peaks, for both the tests. As previously discussed, this condition could be
related to the selected impact section $\tilde{x}_{f}$, which is near to the
nodal points of both the first and the fifth mode shapes $\phi_{1,5}\left(
x_{f}\right)  $. Therefore, in order to get the correct information from the
experimental data in the fitting procedure, we have considered only the
frequency range with maximum coherence, i.e. $30-250$ [Hz] for the beam with
$L=60$ [cm], and $150-500$ [Hz] for the beam with $L=40$ [cm]. In the first
case ($L=60$ [cm]), we have excluded the first peak at $\omega\simeq20$ [Hz],
since it is too near to the zone with low coherence. In the latter case
($L=40$ [cm]), we have excluded both the first ($\omega\simeq50$ [Hz]) and the
second peak ($\omega\simeq120$ [Hz]), because of the heavy drop of the
coherence in correspondence of the first resonance.

The measured frequency response function $H_{m}\left(  \tilde{x},\tilde{x}%
_{f},\omega\right)  $ has been fitted by means of the theoretical FRF
$H_{th}\left(  \tilde{x},\tilde{x}_{f},\omega\right)  $ defined in
Eq.(\ref{FRFTh}), in which only the viscoelastic modulus $E\left(
\omega\right)  $ is unknown. Hence, we have defined the cost function
$\epsilon_{k}$ as the squared difference between the real and imaginary parts
of the theoretical $H_{th}\left(  \tilde{x},\tilde{x}_{f},\omega\right)  $ and
the measured $H_{m}\left(  \tilde{x},\tilde{x}_{f},\omega\right)  $ FRFs:%
\begin{align}
\epsilon_{k}  &  =\sum_{i=n}^{m}\left[  \left(  \operatorname{Re}%
[H_{th}\left(  \tilde{x},\tilde{x}_{f},\omega_{i}\right)  ]-\operatorname{Re}%
[H_{m}\left(  \tilde{x},\tilde{x}_{f},\omega_{i}\right)  ]\right)
^{2}+\right. \label{error}\\
&  \left.  +\left(  \operatorname{Im}[H_{th}\left(  \tilde{x},\tilde{x}%
_{f},\omega_{i}\right)  ]-\operatorname{Im}[H_{m}\left(  \tilde{x},\tilde
{x}_{f},\omega_{i}\right)  ]\right)  ^{2}\right]
\end{align}

The best fit of the theoretical model has been performed by minimizing the
above defined cost function $\epsilon_{k}$, which depends on the number $k$ of
relaxation times considered to characterize the viscoelastic modulus $E\left(
\omega\right)  $. In this manner, the viscoelastic modulus $E\left(
\omega\right)  $ (see definition in the Laplace domain in
Eq.(\ref{ElasticModulusLaplace})), can be determined in terms of i) the
elastic modulus at zero-frequency~$E_{0}$, ii) the relaxation times $\tau_{k}%
$, and iii) the correspondent elastic moduli $E_{k}$. The fundamental novelty
of the presented approach, with respect to the other similar vibration-based
procedures presented in literature (e.g. \cite{Cortes2007}), consists in the
fitting method, which can be optimized by properly choosing the number $k$ of
relaxation times, to correctly fit the beam dynamic response. This number, in
particular, is influenced by the width of the frequency band considered, and
by the amount of damping present in a certain frequency
range.\begin{figure}[ptb]
\begin{center}
\includegraphics[
height=4cm,
]{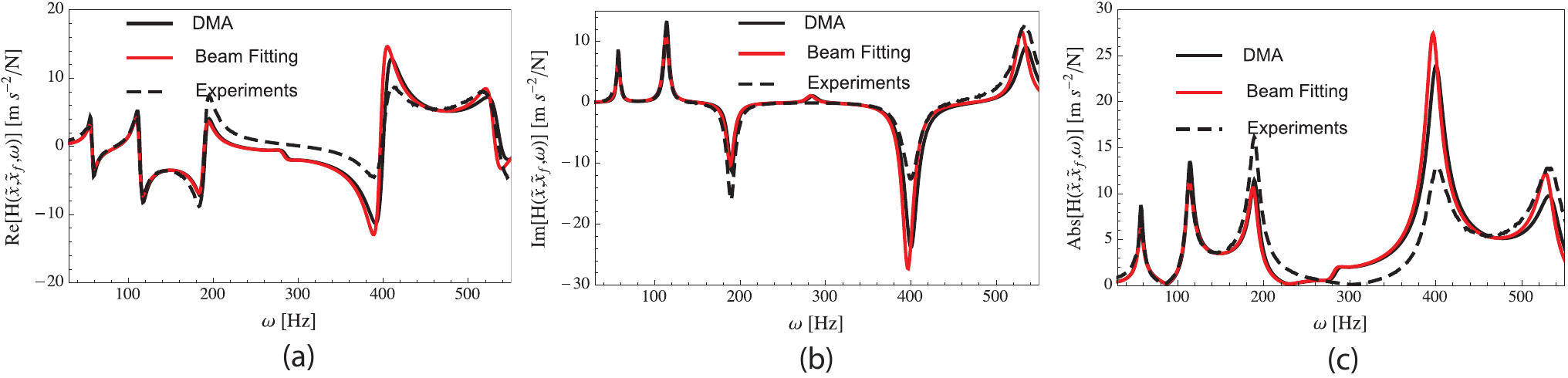}
\end{center}
\caption{The comparison between the measured FRF $H_{m}\left(  \tilde
{x},\tilde{x}_{f},\omega\right)  $ (black dashed lines) and the theoretical
FRF $H_{th}\left(  \tilde{x},\tilde{x}_{f},\omega\right)  $, obtained through
beam-best fitting (red lines) and DMA data (black solid lines), for the beam
with $L=60$ [cm], in the range $30-550$ [Hz], in terms of real part (a),
imaginary part (b) and absolute value (c) of the FRFs.}%
\label{Figure 9}%
\end{figure}

With the aim of assessing the presented technique, we have also experimentally
characterized our viscoelastic material through a Dynamic Mechanical Analyzer
- MCR 702 MultiDrive - Anton Paar GmbH (Tribolab, Politecnico di Bari, Bari,
Italy). However, the DMA approach is different, since the viscoelastic modulus
$E\left(  \omega\right)  $ is directly measured by considering the stress -
strain relation shown in the Eq.(\ref{stress-strain}). Therefore, in order to
define a frequency response function based on DMA results, we need an
analytical form of the viscoelastic modulus $E\left(  \omega\right)  $ to be
considered in Eq.(\ref{FRFTh}). Hence, we have fitted the experimental
viscoelastic modulus $E\left(  \omega\right)  $ measured with DMA, by means of
Eq.(\ref{ElasticModulusLaplace}). In the frequency range $10^{-15}-10^{5}$
[Hz], $50$ relaxation times have been utilized. In Figure \ref{Figure 11} it
is shown the good correlation between the experimental master curve and the
fitted complex modulus, for both the real part (Figure \ref{Figure 11}-a) and
the imaginary part (Figure \ref{Figure 11}-b).

\section{Results and discussions}

The first experimental data set considered is related to the beam with $L=60$
[cm]. From the first iterations, we found that, in order to obtain the best
results, it is preferable to consider two peaks at a time, i.e. the second and
the third resonances in the frequency range $30-150$ [Hz] (see Figure
\ref{Figure 3}-a). The best fitting of the theoretical model (Eq.\ref{error})
has been achieved by means of $11$ relaxation times, and it is shown in Figure
\ref{Figure 4}, where the measured FRF $H_{m}\left(  \tilde{x},\tilde{x}%
_{f},\omega\right)  $ (black dashed lines) is compared with the theoretical
FRFs $H_{th}\left(  \tilde{x},\tilde{x}_{f},\omega\right)  $, obtained by
utilizing Eq.(\ref{FRFTh}), and by considering the viscoelastic modulus
calculated by means of both the beam-fitting procedure (red lines) and the
DMA-fitted data (black solid lines), in terms of real part (a), imaginary part
(b) and absolute value (c) of the FRFs. Interestingly, it is possible to
observe a very good overlapping between the measured curves and the
theoretical FRF obtained with our proposed method. The viscoelastic modulus
$E\left(  \omega\right)  $ calculated by minimizing the cost function
$\epsilon_{k}$ Eq.\ref{error} is shown in Figure \ref{Figure 5} (dashed
lines), where it is compared with the viscoelastic modulus measured with DMA
(solid lines), in the frequency range $30-150$ [Hz].\begin{figure}[ptb]
\begin{center}
\includegraphics[
height=4cm,
]{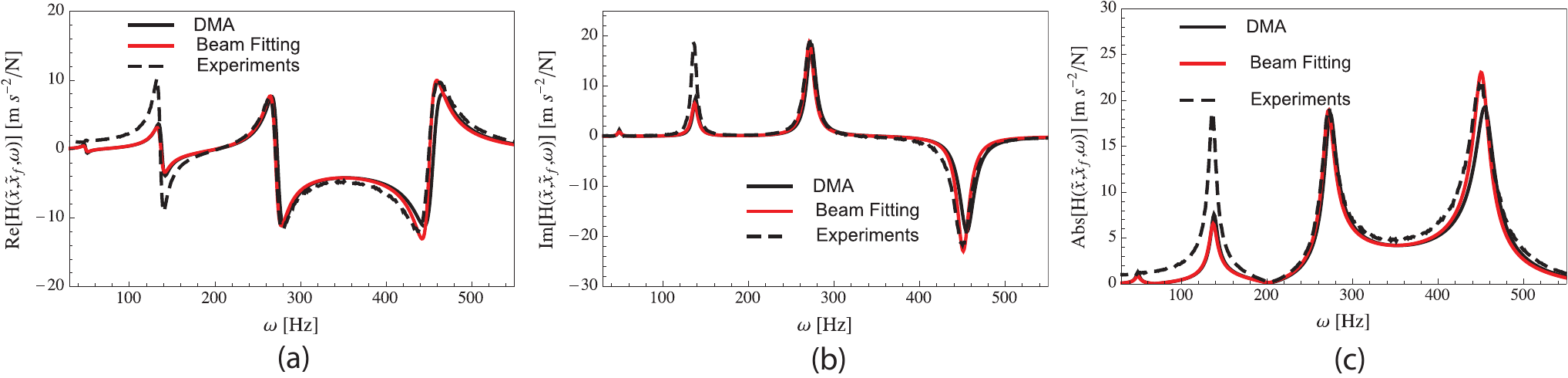}
\end{center}
\caption{The comparison between the measured FRF $H_{m}\left(  \tilde
{x},\tilde{x}_{f},\omega\right)  $ (black dashed lines) and the theoretical
FRF $H_{th}\left(  \tilde{x},\tilde{x}_{f},\omega\right)  $, obtained through
beam-best fitting (red lines) and DMA data (black solid lines), for the beam
with $L=40$ [cm], in the range $30-550$ [Hz], in terms of real part (a),
imaginary part (b) and absolute value (c) of the FRFs.}%
\label{Figure 10}%
\end{figure}

The higher frequency range, i.e. $150-500$ [Hz], has been studied by
investigating the dynamic response of the beam with smaller length, i.e.
$L=40$ [cm]. In this case, we obtained the best fit with $8$ relaxation
times.\ Let us notice that, despite of the broader frequency range now
considered, the number of relaxation times in this case is less than that one
utilized in the previous case (i.e. $11$ relaxation times). The reason is
probably related to the fact that the transition region of
LUBRIFLON$^{\textregistered}$ is found at low frequency (see Figure
\ref{Figure 11}-b), and therefore the more the frequencies are low, the more
the viscoelastic modulus must be characterized through a higher number of
relaxation times to properly describe the fast increase of damping, i.e. of
the imaginary part $\operatorname{Im}[E\left(  \omega\right)  ]$. In Figure
\ref{Figure 6}, we compare the measured FRF $H_{m}\left(  \tilde{x},\tilde
{x}_{f},\omega\right)  $ (black dashed lines) with the theoretical FRFs
$H_{th}\left(  \tilde{x},\tilde{x}_{f},\omega\right)  $ (Eq.(\ref{FRFTh})),
calculated by considering the viscoelastic modulus experimentally obtained by
means of the beam dynamics (red lines) and through DMA (black solid lines), in
terms of real part (a), imaginary part (b) and absolute value (c) of the FRFs.
Also in this case, the proposed approach turns out to be very suitable for the
viscoelastic material characterization. Indeed, it is possible to observe in
Figure \ref{Figure 7} a good overlapping, in the frequency range $150-550$
[Hz], between the viscoelastic moduli $E\left(  \omega\right)  $
experimentally obtained by DMA (solid lines), and by the proposed approach
(dashed lines), both for the real part (Figure \ref{Figure 7}-a) and the
imaginary part (Figure \ref{Figure 7}-b).

At last, the viscoelastic moduli $E\left(  \omega\right)  $, characterized by
means of the vibrational analysis on the beams with lengths $L=60$ [cm]
($30-150$ [Hz]) and $L=40$ [cm] ($150-550$ [Hz]), previously shown
respectively in the Figures \ref{Figure 5}-\ref{Figure 7}, have been fitted
through Eq.(\ref{ElasticModulusLaplace}) in the whole frequency range $30-550$
[Hz]. In Figure \ref{Figure 8}, we show the viscoelastic modulus $E\left(
\omega\right)  $ determined by means of the beam dynamics (dashed lines) and
the one measured with DMA (solid lines). For both the real part (Figure
\ref{Figure 8}-a) and the imaginary part (Figure \ref{Figure 8}-b) of the
viscoelastic modulus $E\left(  \omega\right)  $, we obtained a fine matching,
thus finally assessing the method proposed in this paper. In Figure
\ref{Figure 9}, we compare, in the range $30-550$ [Hz], the measured FRF
$H_{m}\left(  \tilde{x},\tilde{x}_{f},\omega\right)  $ (black dashed lines)
with the theoretical curves $H_{th}\left(  \tilde{x},\tilde{x}_{f}%
,\omega\right)  $, obtained through the so calculated viscoelastic modulus
$E\left(  \omega\right)  $ (Figure \ref{Figure 8}) (red lines) and by means of
DMA data (black solid lines), for the beam with $L=60$ [cm]. It is important
to highlight that the two theoretical FRFs $H_{th}\left(  \tilde{x},\tilde
{x}_{f},\omega\right)  $ are overlapped in all the frequency range, while the
experimental FRF follows the theoretical curves only in the range $30-150$
[Hz], where we obtained maximum coherence (see Figure \ref{Figure 3}-a). The
"non-dectected peak" at $300$ [Hz] in the experimental acquisitions, is
strictly related to the drop in the coherence function, and could be the
origin of the non perfect overlapping between the theoretical and experimental
FRFs in the range $150-550$ [Hz].

Similar reasonings can be made for the results obtained from the beam with
$L=40$ [cm]. In Figure \ref{Figure 10} we report the theoretical and the
experimental FRFs in this case, where it is evident the good correspondence
between the theoretical functions $H_{th}\left(  \tilde{x},\tilde{x}%
_{f},\omega\right)  $, obtained through our proposed procedure (red lines) and
by means of DMA data (black solid lines), in all the frequency range $30-550$
[Hz]. However, also in this case, the experimental curve follows the
theoretical ones in a limited range, i.e. $200-550$ [Hz], which is far from
the presence of a "non-dectected peak" at around $65$ [Hz] (see Figure
\ref{Figure 3}-b), that probably caused a drop in the coherence function.
Moreover, at very small frequencies ($\sim10$ [Hz]), in both the experimental
acquisitions (Figure \ref{Figure 3}-a,b) it should be observed that coherence
tends to decrease below limit values, i.e. $<0.8$, because of the intrinsic
problematic of the instrumentations, especially of the impact hammer.

In light of what has emerged from the results shown so far, some remarks
should be made, in order to define guidelines for the procedure proposed in
this paper. First, we found that for a good fitting of the vibrational
response of the beam, the frequency range where coherence is not maximum, as
well as some peaks near these areas, should be excluded from the fitting
calculations. Furthermore, it has been shown that, the more we proceed towards
frequencies where damping is high, i.e. versus the transition zone of the
viscoelastic material, the more we need to consider a narrow frequency band to
fit the beam response, and an increasing number of relaxation times to
describe the viscoelastic modulus $E\left(  \omega\right)  $ is required too.

In conclusion, the proposed method for the characterization of the
viscoelastic materials, has revealed to be very efficient, easy to use, and
reliable with inexpensive instrumentation. Moreover, the analytical model here
presented and used to fit the experimental response of the beam, in
particular, has proven to be accurate. The comparison between the viscoelastic
moduli $E\left(  \omega\right)  $ characterized by means of our technique and
through DMA, indeed, finally assessed the possibility to retrieve this so
important mechanical quantity, by simply investigating the dynamics of a
viscoelastic beam. We also found that, the idea to consider more beams with
different lengths, is very useful to increase the frequency range of interest,
and, in principle, by studying the dynamics of even longer or shorter beams,
it is possible to cover wider frequency ranges. However, it should be
highlighted that in order to obtain a range comparable with the one usually
covered by DMA, a different instrumentation should be utilized. In particular,
the impact hammer represents a limit in this direction, and it should be
substituted with an electrodynamic shaker, which enables to investigate a
wider frequency range maintaining high coherence. At last, also by controlling
the surrounding temperature, is possible to enlarge the range of interest,
i.e. by a frequency shift of the viscoelastic modulus $E\left(  \omega\right)
$ under study. These last two modifications to the actual experimental setup,
will be object of further investigations.

\subsection{Conclusions}

In this paper we have presented a very simple and accurate experimental
approach for determining the complex modulus of viscoelastic materials. By
means of the vibrational behaviour of suspended viscoelastic beams with
different lengths, we have characterized the complex modulus of
LUBRIFLON$^{\textregistered}$ by fitting the measured response through an
accurate analytical model of the beam dynamics, which takes into account
multiple relaxation times of the material. In particular, the possibility to
properly select the number of relaxation times in a frequency range of
interest, turned out to be a key factor to obtain very good results. The
instrumentation utilized in our experiments is inexpensive and easy to use,
and it consists of an impact hammer and a suspended beam, instrumented by
means of an accelerometer connected to a data acquiring module. Comparisons
with DMA measurements demonstrate the validity of the proposed technique on a
frequency range which could be comparable with the one usually covered by DMA
technique. In conclusion, the proposed procedure represents a valid
alternative approach to DMA, and can be considered as a significative step
forward the improvement of the mechanical characterization of viscoelastic materials.

\end{document}